\documentclass[twoside,12pt]{report}
\usepackage{graphicx}
\usepackage{amssymb, amsmath, amsxtra}
\usepackage{epsfig,subfigure}

\usepackage{latexsym}
\usepackage{bm}

 \usepackage{ntheorem}
           \theoremstyle{plain}
                      {\theorembodyfont{\rmfamily}
                      \theoremseparator{.}

           \theoremstyle{plain}
           
           \theoremstyle{plain} 
           \theoremstyle{plain}

           \theoremstyle{plain}
           \newtheorem{rem}{Remark}[section]
            \newtheorem{prop}{Proposition}[section]}

\begin{document}

\begin{center}
{\Large \textbf{Kadomtsev-Petviashvili equation:\\ Nonlinear
self-adjointness\\[.5ex] and
  conservation laws}}\\[2ex]
 Nail H. Ibragimov\\
 Department of Mathematics and Science, Blekinge Institute
 of Technology,\\ 371 79 Karlskrona, Sweden
 \end{center}

 \noindent
\textbf{Abstract.} The method of nonlinear self-adjointness is
applied to the
 Kadomtsev-Petviashvili equation. The infinite set of conservation laws associated with
 the infinite algebra of Lie point symmetry of the KP equation is constructed.\\[2ex]
 \hfill

 \begin{table}[b]
  \begin{tabular}{l}\hline
 \copyright ~2011 N.H. Ibragimov.\\ Published in \textit{Archives of ALGA}, vol. 7/8, 2010-2011, pp. 139-146.\\
  \end{tabular}
 \end{table}

 \tableofcontents
  \newpage

 \section{Lie point symmetries of the KP equation}
 \label{KP}
 \setcounter{equation}{0}


  It is convenient for our purposes to write the Kadomtsev-Petviashvili \cite{KP70} equation
 \begin{equation}
 \label{KP.eq1}
 D_x(u_t - u u_x - u_{xxx}) = u_{yy},
 \end{equation}
 or
 $$
u_{tx} - u u_{xx} - u^2_x - u_{xxxx} = u_{yy},
 $$
 in the form of the system (see \cite{nov80} and the references
 therein)
 \begin{equation}
 \label{KP.eq2}
 u_t - u u_x - u_{xxx} - \omega_y = 0, \quad \omega_x - u_y = 0.
 \end{equation}

 The Lie algebra of the infinitesimal symmetries of the KP equation (\ref{KP.eq1}) as well as of the system (\ref{KP.eq2})
 is quite similar to the infinite-dimensional symmetry Lie algebra of the Lin-Reissner-Tsien
 equation \cite{lin-rei-tsi48}
  \begin{equation}
 \label{KP.eq3}
 2 \varphi_{tx} + \varphi_x \varphi_{xx} = \varphi_{yy},
  \end{equation}
 describing the non-steady state
 gas flow with transonic velocities. The symmetry Lie algebra of Eq. (\ref{KP.eq3})
 has been obtained in \cite{mam69} (see also \cite{ibr72}, \S 28). It involves five arbitrary functions of $t$
 and contains, in particular
 the  following operators:
 \begin{equation}
 \label{KP.eq4}
 \begin{split}
 X_f & =3 f(t)\frac{\partial}{\partial t} + (f'(t)x+f''(t)y^2)\frac{\partial}{\partial
 x} +2f'(t)y \frac{\partial}{\partial y}\\[1.5ex]
 & +\big[f''(t)x^2+2f'''(t)xy^2+\frac13
 f^{(4)}(t)y^4-f'(t)\varphi\big]\frac{\partial}{\partial
 \varphi}\,,
 \end{split}
  \end{equation}
 \begin{equation}
 \label{KP.eq5}
  X_g =g(t)\frac{\partial}{\partial y} + g'(t)y \frac{\partial}{\partial x}
  +\big[2g''(t)xy+\frac23 g'''(t)y^3\big]\frac{\partial}{\partial \varphi}\,,
 \end{equation}
 \begin{equation}
 \label{KP.eq6}
 X_h = h(t) \frac{\partial}{\partial x} +\big[2 h'(t)x
 + 2h''(t) y^2\big]\frac{\partial}{\partial \varphi}\,\cdot
 \end{equation}
 The system (\ref{KP.eq2}) admits
 the following operators (compare with (\ref{KP.eq4})-(\ref{KP.eq6})):
 \begin{equation}
 \label{KP.eq7}
 \begin{split}
 & X_f =3 f\frac{\partial}{\partial t}
  + (f'x+\frac{1}{2}\,f''y^2)\frac{\partial}{\partial x}
  + 2f'y \frac{\partial}{\partial y}\\[1.5ex]
  &-\big[2 f'u + f''x + \frac{1}{2}\,f'''y^2\big]\frac{\partial}{\partial u}
   - \big[3 f' \omega + f'' y u + f''' xy
  +\frac{1}{6} f^{(4)}y^3\big]\frac{\partial}{\partial \omega}\,,
 \end{split}
  \end{equation}
 \begin{equation}
 \label{KP.eq8}
  X_g = 2 g\frac{\partial}{\partial y} + g'y \frac{\partial}{\partial x}
   - g''y \frac{\partial}{\partial u} -
   \big[g' u + g''x+\frac{1}{2}\, g'''y^2\big]\frac{\partial}{\partial \omega}\,,
 \end{equation}
 \begin{equation}
 \label{KP.eq9}
 X_h = h \frac{\partial}{\partial x} - h' \frac{\partial}{\partial u}
 - h'' y \frac{\partial}{\partial \omega}\,,
 \end{equation}
 where $f, g, h$ are three arbitrary functions of $t.$
 We will ignore the obvious symmetry
 $$
 X_\alpha = \alpha(t) \frac{\partial}{\partial  \omega}
 $$
 of the system (\ref{KP.eq2}) describing the addition to $\omega$ an
 arbitrary function of $t.$

 Note, that the operators (\ref{KP.eq7})-(\ref{KP.eq9}) considered without the
 term $\frac{\partial}{\partial \omega}$ span the infinite-dimensional Lie algebra of
 symmetries of the KP equation (\ref{KP.eq1}). They coincide (up to normalizing coefficients)
 with the symmetries of the KP equation that were first
 obtained by F. Schwarz in 1982 (see also \cite{dklw85} and
 \cite{ame-rog89}).

 \section{Nonlinear self-adjointness}
 \label{KP:2}
 \setcounter{equation}{0}

 The Kadomtsev-Petviashvili equation written in the form (\ref{KP.eq1}) or in the form of
 the system (\ref{KP.eq2}) does not have a Lagrangian. Let us investigate
 the KP equation for nonlinear self-adjointness \cite{ibr10a}.

 The formal Lagrangian for the system (\ref{KP.eq2}) is written
 \begin{equation}
 \label{KP.eq10}
 {\cal L} = v (u_t - u u_x - u_{xxx} - \omega_y) + z(\omega_x -
 u_y).
  \end{equation}
  The reckoning shows that
  \begin{align}
 & \frac{\delta {\cal L}}{\delta u} = - v_t + u v_x + v_{xxx} +
 z_y,\notag\\[1ex]
  & \frac{\delta {\cal L}}{\delta \omega} = v_y - z_x.\notag
  \end{align}
 Hence we can write the adjoint system to (\ref{KP.eq2}) in the form
 \begin{equation}
 \label{KP.eq11}
 v_t - u v_x - v_{xxx} - z_y = 0, \quad z_x - v_y = 0.
 \end{equation}
 Eqs. (\ref{KP.eq11}) become identical with the KP
 equations (\ref{KP.eq2}) upon the substitution
 \begin{equation}
 \label{KP.eq12}
 v = u, \quad z = \omega.
 \end{equation}
 It means that the system (\ref{KP.eq2}) is nonlinearly self-adjoint, specifically it is
 quasi self-adjoint (\cite{ibr10a}, Section 1.6, Definition 1.3).

 \section{Conservation laws provided by Lie point symmetries}
 \label{KP:3}
 \setcounter{equation}{0}

 Let us introduce the notation
 $$
 x^1 = t, \quad x^2 = x, \quad x^3 = y, \quad u^1 = u, \quad u^2 =
 \omega
 $$
 and write conservation laws in the form of the differential
 equation
 \begin{equation}
 \label{KP.eq13}
 \left[D_t(C^1) + D_x(C^2) + D_y(C^3)\right]_{(\ref{KP.eq2})} =0,
 \end{equation}
 where $|_{(\ref{KP.eq2})}$ means that the equation holds on the solutions of
the system (\ref{KP.eq2}).

We will use the general formula given in \cite{{ibr10a}} for
constructing the conserved vector associated with symmetry
$$
X = \xi^i \frac{\partial}{\partial x^i} + \eta^\alpha
\frac{\partial}{\partial u^\alpha}
$$
of a system of differential equations with a classical or formal
Lagrangian ${\cal L}.$ Since the maximum order of derivatives
involve in formal Lagrangian ${\cal L}$ given by Eq. (\ref{KP.eq10})
is equal to three, this formula is written
 \begin{align}
 C^i & = \xi^i {\cal L}+W^\alpha
 \left[\frac{\partial {\cal L}}{\partial u_i^\alpha} -
 D_j \left(\frac{\partial {\cal L}}{\partial u_{ij}^\alpha}\right)
 + D_j D_k\left(\frac{\partial {\cal L}}{\partial u_{ijk}^\alpha}\right)\right]\notag\\[1.5ex]
 &+D_j\left(W^\alpha\right)
 \left[\frac{\partial {\cal L}}{\partial u_{ij}^\alpha} -
 D_k \left(\frac{\partial {\cal L}}{\partial
 u_{ijk}^\alpha}\right)\right]
 + D_j D_k\left(W^\alpha\right) \frac{\partial {\cal L}}{\partial u_{ijk}^\alpha}\,,\notag
 \end{align}
where
 $$
 W^\alpha = \eta^\alpha - \xi^j u_j^\alpha.
 $$

 We will apply the above formula to the symmetries (\ref{KP.eq7})-(\ref{KP.eq9}). Invoking
 that the system (\ref{KP.eq2}) is nonlinearly self-adjoint with the substitution (\ref{KP.eq12}), we will
 replace in $C^i$ the \textit{non-physical variables} $v$ and $z$
 with $u$ and $\omega,$ respectively, thus arriving to conserved
 vectors for the KP system.
 Since the formal Lagrangian (\ref{KP.eq10}) vanishes
 on the solutions of the system (\ref{KP.eq2}), we can omit the term $\xi^i {\cal L}$
 and take the formula for the conserved vector in the
 following form:
  \begin{align}
 \label{KP.eq14}
 C^i & = W^\alpha\,
 \Big[\frac{\partial {\cal L}}{\partial u_i^\alpha} -
 D_j \Big(\frac{\partial {\cal L}}{\partial u_{ij}^\alpha}\Big)
 + D_j D_k\Big(\frac{\partial {\cal L}}{\partial u_{ijk}^\alpha}\Big)\Big]\\[1.5ex]
 &+D_j\big(W^\alpha\big)\,
 \Big[\frac{\partial {\cal L}}{\partial u_{ij}^\alpha} -
 D_k \Big(\frac{\partial {\cal L}}{\partial
 u_{ijk}^\alpha}\Big)\Big]
 + D_j D_k\big(W^\alpha\big)\frac{\partial {\cal L}}{\partial u_{ijk}^\alpha}\,,\notag
 \end{align}
 where
 \begin{equation}
 \label{KP.eq15}
  W^\alpha = \eta^\alpha - \xi^j u_j^\alpha, \quad \alpha = 1, 2.
 \end{equation}

 Using in (\ref{KP.eq14}) the expression (\ref{KP.eq10}) for
 ${\cal L}$ and eliminating $v$ and $z$ by mans of Eqs. (\ref{KP.eq12})
 we obtain
 \begin{equation}
 \label{KP.eq16}
 \begin{split}
 & C^1 = u W^1,\\[1ex]
 & C^2 = - (u^2 + u_{xx}) W^1 + \omega W^2 + u_x D_x(W^1) - u D_x^2(W^1),\\[1ex]
 & C^3 = - \omega W^1 - u W^2.
 \end{split}
 \end{equation}

 The expressions (\ref{KP.eq15}) for the operator (\ref{KP.eq7}) are
 written:
 \begin{align}
 \label{KP.eq17}
 & W^1 = - 3 f u_t - (2 u + x u_x + 2 y u_y) f' - \big(x + \frac{1}{2}\,y^2
  u_x\big) f'' - \frac{1}{2}\,y^2 f''',\\[1ex]
 & W^2 = - 3 f \omega_t - (3 \omega + x \omega_x + 2 y \omega_y) f' - \big(y u + \frac{1}{2}\,y^2
  \omega_x\big) f'' - x y f''' - \frac{1}{6}\,y^3 f^{(4)}.\notag
 \end{align}

 Substituting $W^1$ given by (\ref{KP.eq17}) in the first equation (\ref{KP.eq16}) and eliminating $u_t$ by using
 the first equation (\ref{KP.eq2}) we obtain
 \begin{equation}
 \label{KP.eq18}
 \begin{split}
  C^1  = & - 3 (u^2 u_x + u u_{xxx} + u \omega_y)f - (2 u^2 + x u u_x + 2 y u u_y) f'\\[1ex]
 &  - \big(x u + \frac{1}{2}\,y^2 u u_x\big) f'' - \frac{1}{2}\,y^2 u f'''.
 \end{split}
 \end{equation}
 Now we  single out the
 total derivatives with respect to $x$ and $y,$ by taking into account the second equation
  (\ref{KP.eq2}), and rewrite (\ref{KP.eq18}) in the form
 \begin{equation}
 \label{KP.eq19}
  C^1  =  - \frac{1}{2}\, f' u^2 - \big(x f'' + \frac{1}{2}\,y^2 f'''\big) u
  + D_x (P) + D_y (Q),
 \end{equation}
 where
 \begin{align}
 \label{KP.eq20}
  &P =  \bigg[\frac{3}{2}\,u^2_x
 + \frac{3}{2}\,\omega^2 - u^3 - 3 u u_{xx}\bigg]f - \frac{1}{2}\, f' x u^2
 - \frac{1}{4}\, f'' y^2 u^2,\notag  \\[1ex]
 &  Q = -  3 f u \omega - f' y u^2.
 \end{align}
 Thus, the first component of the conserved  vector can be reduced
 to the form
 \begin{equation}
 \label{KP.eq21}
  \widetilde C^1  =  - \frac{1}{2}\, f' u^2 -
  \big(x f'' + \frac{1}{2}\,y^2 f'''\big) u.
 \end{equation}
 \begin{rem}
 \label{KP:rem1}
 In reducing (\ref{KP.eq18}) to the form (\ref{KP.eq19})
 we use simple identities such as
 \begin{align}
 & u u_{xxx} = D_x(u u_{xx}) - u_x u_{xx} = D_x\Big(u u_{xx} -
 \frac{1}{2}\, u_x^2\Big),\notag\\[1ex]
 & u \omega_y = D_y(u \omega) - \omega u_y = D_y(u \omega) - \omega \omega_x
  = D_y(u \omega) - D_x\Big(\frac{1}{2}\, \omega ^2\Big),\notag\\[1ex]
 & 2 y u u_y = D_y(y u^2) -  u^2.\notag
 \end{align}
 \end{rem}

 To find the second component of the conserved vector, we substitute
 the expressions (\ref{KP.eq17}) of $W^1, W^2$ in the second
  equation (\ref{KP.eq16}), add $D_t (P)$  with $P$ defined in
  (\ref{KP.eq20}) and obtain:
 \begin{align}
 \bar C^2 & = C^2 + D_t(P) \notag\\[1ex]
 &  = \big(u u_{xx} + \frac{1}{3}\,u^3 - \frac{1}{2}\,u_x^2 - \frac{1}{2}\, \omega^2\big) f'
 + \big(x u_{xx} + \frac{1}{2}\, x u^2 - u_x\big) f''\notag\\[1ex]
 & + \frac{1}{4}\, (y^2 u^2 + 2 y^2 u_{xx} - 4 x y \omega) f'''
 - \frac{1}{6}\, y^3 \omega f^{(4)} + D_y(R),\notag
 \end{align}
 where
 \begin{equation}
 \label{KP.eq22}
  R = \big(2 y u u_{xx} + \frac{2}{3}\, y u^3 - y u_x^2 - y \omega^2 - x u \omega \big) f'
  - \frac{1}{2}\, y^2 u \omega f''.
 \end{equation}
 Thus, the second component of the conserved can be reduced to
  \begin{align}
  \label{KP.eq23}
 \widetilde C^2 & = \big(u u_{xx} + \frac{1}{3}\,u^3 - \frac{1}{2}\,u_x^2 - \frac{1}{2}\, \omega^2\big) f'
 + \big(x u_{xx} + \frac{1}{2}\, x u^2 - u_x\big) f''\notag\\[1ex]
 & + \frac{1}{4}\, (y^2 u^2 + 2 y^2 u_{xx} - 4 x y \omega) f'''
 - \frac{1}{6}\, y^3 \omega f^{(4)}.
 \end{align}

 Finally, the third component of the conserved vector is obtained by
 substituting the expressions (\ref{KP.eq17}) of $W^1, W^2$ in the third
  equation (\ref{KP.eq16}) and adding $D_t (Q), D_x(R)$  with $Q, R$ defined in
  (\ref{KP.eq20}), (\ref{KP.eq20}):
 $$
 \widetilde C^3 = C^3 + D_t(Q) + D_x(R).
 $$
   The reckoning yields:
 \begin{equation}
 \label{KP.eq24}
 \widetilde C^3 = u \omega f' + x \omega f''
 +  \big(x y u + \frac{1}{2}\, y^2\omega \big) f'''
  + \frac{1}{6}\, y^3 u f^{(4)}.
 \end{equation}

 Ignoring the tilde in the quantities (\ref{KP.eq21}), (\ref{KP.eq23}), (\ref{KP.eq24}),
 we summarize the result in the following statement.
 \begin{prop}
 \label{Prop1}
 The infinitesimal symmetry (\ref{KP.eq7}) of the
 Kadomtsev-Petviashvili equations (\ref{KP.eq1})
 provides the conserved vector $C = (C^1, C^2, C^3)$ with the components
 \begin{equation}
 \label{KP.eq25}
 \begin{split}
 C^1 & =  - \frac{1}{2}\, f' u^2 -
  \big(x f'' + \frac{1}{2}\,y^2 f'''\big) u,\\[1.5ex]
 C^2 & = \big(u u_{xx} + \frac{1}{3}\,u^3 - \frac{1}{2}\,u_x^2 - \frac{1}{2}\, \omega^2\big) f'
 + \big(x u_{xx} + \frac{1}{2}\, x u^2 - u_x\big) f''\\[1ex]
 & + \frac{1}{4}\, (y^2 u^2 + 2 y^2 u_{xx} - 4 x y \omega) f'''
 - \frac{1}{6}\, y^3 \omega f^{(4)},\\[1.5ex]
 C^3 & = u \omega f' + x \omega f''
 +  \big(x y u + \frac{1}{2}\, y^2\omega \big) f'''
  + \frac{1}{6}\, y^3 u f^{(4)}.
  \end{split}
 \end{equation}
 \end{prop}

 \begin{rem}
 \label{KP:rem2}
 The validity of the conservation equation (\ref{KP.eq13}) for the vector (\ref{KP.eq25})
 follows from the following equation:
 \begin{equation}
 \label{KP.eq26}
 \begin{split}
 & D_t(C^1) + D_x(C^2) + D_y(C^3)\\[1ex] & = (u f' + x f'' + \frac{1}{2}\, y^2f''')
 (u_{xxx} + u u_x + \omega_y - u_t)\\[1ex]
 & + (\omega f' + xy f''' + \frac{1}{6}\, y^3 f^{(4)})
 (u_y - \omega_x).
  \end{split}
 \end{equation}
 \end{rem}

 The similar calculations  for the operators (\ref{KP.eq8}) and
 (\ref{KP.eq9}) yield the following.
 \begin{prop}
 \label{Prop2}
 The symmetry (\ref{KP.eq8}) of
 the system (\ref{KP.eq1})
 provides the conserved vector $C = (C^1, C^2, C^3)$ with the components
 \begin{equation}
 \label{KP.eq27}
 \begin{split}
 C^1 & =  y u g'',\\[1.5ex]
 C^2 & = \left(x \omega - y u_{xx} - \frac{1}{2}\,y u^2\right) g''
 + \frac{1}{2}\, y^2 \omega g''',\\[1ex]
 C^3 & = - (x u + y \omega) g'' - \frac{1}{2}\, y^2 u g'''.
  \end{split}
 \end{equation}
 \end{prop}
 \begin{prop}
 \label{Prop3}
 The symmetry (\ref{KP.eq9}) of
 the system (\ref{KP.eq1})
 provides the conserved vector $C = (C^1, C^2, C^3)$ with the components
 \begin{equation}
 \label{KP.eq28}
 \begin{split}
 C^1 & =  u h',\\[1.5ex]
 C^2 & = y \omega h'' - \left(u_{xx} + \frac{1}{2}\, u^2\right) h',\\[1ex]
 C^3 & = - \omega h' - y u h''.
  \end{split}
 \end{equation}
 \end{prop}

Different approaches to construction of conservation laws for the KP
 equation can be found, e.g. in \cite{zak-shu79, oev-fuc81, inf-fry83}.
 In particular, the Lagrangian approach and the Noether theorem are used in the paper \cite{inf-fry83}
 which contains an interesting discussion of the infinite set of conservation laws.
 Note that the second equation of the system (\ref{KP.eq2})
 guarantees that the vector field $(u, \omega)$ has the potential
 $\phi$ defined by $u = \phi_x, \ \omega =\phi_y.$ Then the system (\ref{KP.eq2}) is
 replaced by the \textit{potential KP equation}
 \begin{equation}
 \label{KP.eq29}
 \phi_{xt} - \phi_x \phi_{xx} - \phi_{xxxx} - \phi_{yy} = 0
 \end{equation}
 which, unlike equation (\ref{KP.eq1}) or the system (\ref{KP.eq2}), has a
 Lagrangian, namely
 \begin{equation}
 \label{KP.eq30}
  L = - \frac{1}{2}\, \phi_x \phi_t  + \frac{1}{6}\, \phi_x^3 + \frac{1}{2}\, \phi_y^2
  - \frac{1}{2}\, \phi_{xx}^2\,.
 \end{equation}
  Now the Noether theorem can be used upon rewriting the symmetries
  of the KP equation in terms of the potential $\phi.$ This approach
  is used in the paper \cite{rosh06} which contains profound results
  on the conservation  laws associated with the infinite algebra of Lie
  point symmetries of Eq. (\ref{KP.eq29}). In particular, it is
  demonstrated their that the differential and integral forms of the
  conservation laws are equivalent only when the functions $f(t), g(t), h(t)$
 in the symmetries (\ref{KP.eq7})-(\ref{KP.eq9}) are low-order
 polynomials. For the details I refer the reader to \cite{rosh06}.


 \end{document}